\documentclass{article}
\usepackage{spconf,amsmath,graphicx}
\usepackage{verbatim}

\usepackage{multirow}
\usepackage{hyperref}
\usepackage{booktabs}

\title{Learning to detect novel and fine-grained acoustic sequences \\ using pretrained audio representations}

\name{Vasudha Kowtha, Miquel Espi Marques, Jonathan Huang, Yichi Zhang, Carlos Avendano}
\address{Apple}

\begin{document}

\maketitle

\begin{abstract}
This work investigates pretrained audio representations for few shot Sound Event Detection. We specifically address the task of few shot detection of novel acoustic sequences, or sound events with semantically meaningful temporal structure, without assuming access to non-target audio. We develop procedures for pretraining suitable representations, and methods which transfer them to our few shot learning scenario. Our experiments evaluate the general purpose utility of our pretrained representations on AudioSet, and the utility of proposed few shot methods via tasks constructed from real-world acoustic sequences. Our pretrained embeddings are suitable to the proposed task, and enable multiple aspects of our few shot framework.  
\end{abstract}


\section{Introduction}
\label{sec:intro}
\vspace{-2mm}

\begin{figure}[!t]
\begin{minipage}[b]{1.0\linewidth}
  \centering
  \centerline{\includegraphics[width=7.9cm]{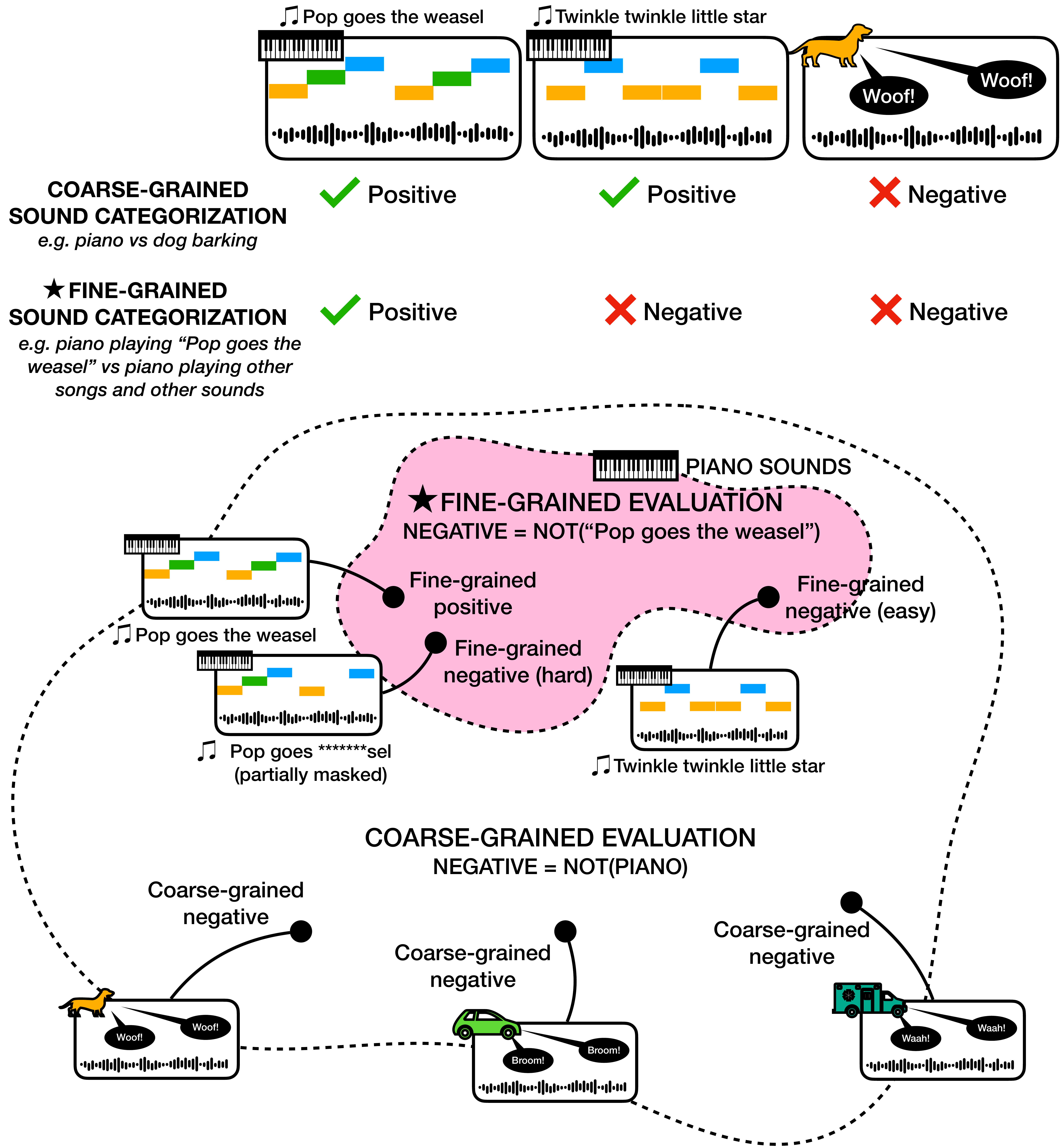}}
\end{minipage}
\vspace{-8mm}
\caption{Acoustic Sequences \& Fine-Grained Evaluation}
\vspace{-8mm}
\label{fig:fig1}
\end{figure}

In the Sound Event Detection (SED) task, the Few Shot Learning (FSL) paradigm suggests a path towards sample-efficient specification and detection of individually desired and idiosyncratic sound categories, which depart from standard labeled audio ontologies. In this work, we explore FSL for the detection of \textit{acoustic sequences}, such as the musical phrase ``pop-goes-the-weasel''. Sequences of this form have distinct temporal structure: if this musical phrase was significantly occluded or scrambled in time, the class identity would change (Fig. \protect\ref{fig:fig1}). The majority of prior work in FSL-SED concerns the detection or classification of coarser-grained sound categories, where \textit{specific sequences} such as ``pop-goes-the-weasel'' or ``twinkle-twinkle-little-star'' might both be considered members of the same coarser grained class of ``ring tone'' or ``music''.  In this typical evaluation setting, FSL approaches are evaluated on their ability to distinguish coarse-grained target sound events from non-target sound classes, which are often significantly dissimilar from the target class. In contrast, we assess the robustness of our proposed approach to detecting specific sequences in a challenging few shot \textit{fine-grained evaluation setting}, wherein all target and non-target sound events belong to the same coarse-grained category.

\begin{figure}[!t]
\begin{minipage}[b]{1.0\linewidth}
  \centering
  \centerline{\includegraphics[width=1.0\linewidth]{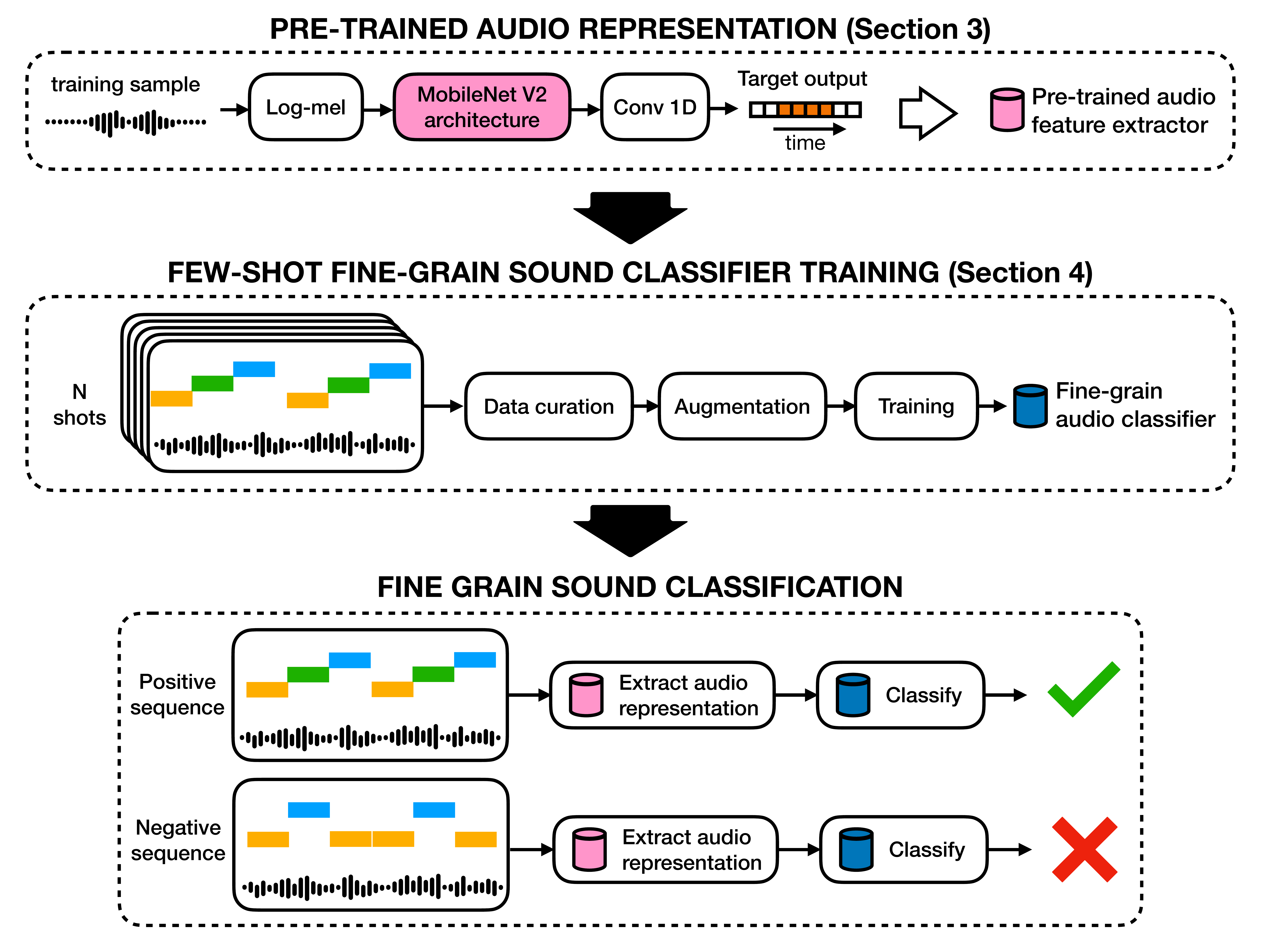}}
\end{minipage}
\vspace{-5mm}
\caption{Training \& Inference Overview} \label{fig:fig2}
\vspace{-5mm}
\end{figure}

We approach this task by jointly designing the representation space of a suitable pretrained backbone embedding extractor, as well as the components of a few shot learning approach which leverages these representations. All in a unified framework (Fig. \protect\ref{fig:fig2}) which: (1) does not require specialized or additional data for training the embedding, and (2) does not require negative data for FSL. We believe this is novel, compared with typical approaches that investigate FSL systems for generic and weak labeled sound event representations. 

\textbf{Our Contribution}. We demonstrate (1) methods for effectively pretraining and distilling CNN-based baseline classifiers for the generic SED task, with competitive benchmarking performance on AudioSet; (2) practical FSL-SED evaluation contexts where designing pretrained representations with improved temporal resolution outperforms 
weak-label-training SED embeddings, and a method for pretraining such representations; and (3) an FSL framework for learning fine-grained binary classifiers which are pretrained from weakly-labeled audio without assuming access to non-target audio samples, and evaluated in a challenging fine-grained setting to detect specific acoustic sequences. 

\section{Related Work}

Few shot learning of sound events has been previously studied. For example, specific contexts such as rare sounds \protect\cite{shimada_icassp_2020_rare_background_noise, zhao_2022_arxiv_rare_sed_adaptive_mask_augment}, transient sounds \protect\cite{chou_icassp_2019_transients_matching}, vocal sounds \protect\cite{salamon_icassp_2020_fewshot_sed} have been explored. This work proposes approaches for detecting acoustic sequences (especially longer sequences e.g. $>$ 5 sec.), which is relatively unexplored in this literature. 

While recent work in the image domain has emphasized the importance of leveraging an appropriate pretrained representation space \protect\cite{chen_2019_few_shot_review, tenenbaum_2020_embedding, liu_2020_embarassing_baseline}, such a shift is yet to be seen in the few shot sound detection literature. Whether this is due to advancements in embedding pretraining in the image domain, or fragility of meta-learning methods to aspects of distribution shift \protect\cite{chen_2019_few_shot_review}

is unclear. It has been suggested that audio-specific considerations, such as polyphony, acoustic degradations and background noise, and weakly labeled data may contribute to a lack of one-size-fits-all approach across both the audio and visual domains \protect\cite{salamon_2021_rethinking_fewshot}. However, previous studies have not investigated methods which co-design both representation space alongside few-shot algorithm components, with attentive handling of aspects such as the lack of availability of non-target audio (negatives). Thus, we investigate the potential of carefully chosen pretrained representations, evaluated under robustness to domain shift, alongside pragmatic handling of audio-specific considerations such as reverberation, weakly labeled targets and aspects of polyphony.

\section{Acoustic Representation Pretraining}
\label{sec:representation}

We build on standard pretrained multi-class sound event classifiers to create a flexible representation capable of fine-grained acoustic sequence discrimination.

\textbf{Training Data}. Several notable datasets have been developed 
for sound classification \protect\cite{piczak2015esc}, \protect\cite{chen2020vggsound}, \protect\cite{gemmeke2017audio}, \protect\cite{fonseca2021fsd50k}.  
For FSL, it is desirable to have a representation that covers a great variety of sounds; for training a robust representation, a large quantity of data is important.  We chose AudioSet \protect\cite{gemmeke2017audio}, with $527$ categories of sounds,  approximately 2 million training, and 20k evaluation examples. Typical sound classification datasets, including AudioSet, are \textbf{weak labeled}: a label is tagged if the sound event for that class occurs anywhere in a file. It would be preferable to use frame-level targets instead, to train models capable of representing temporal dynamics in the audio. While access to \textbf{strong labels} might enable this, via annotated start \& end times per sound event, the cost of strong-labeling for data at the scale of AudioSet is prohibitive.

\vspace{-2mm}
\subsection{Model architectures}
\label{subsec:architectures}
\vspace{-2mm}

We train three ConvNet architectures: 1) a reference model which is trained to maximize performance on weak labeled AudioSet training targets, 2) a smaller MobileNetV2 model, of a size suitable for running on a mobile device, distilled from a larger teacher model, and also trained using the weak labeled targets, and 3) a model with the same ConvNet architecture as 2, but with an output layer suitable for generating estimated strong label predictions. These architectures all support flexible input duration at inference time. The input to the ConvNets are 64-dimension logmel spectrograms extracted with 25 ms input frames at 10 ms hops.

\textbf{Reference Model}. We use a ResNet-50 \protect\cite{he2016deep} backbone for maximizing the sound classification performance.  The output of the ResNet is pooled channel-wise, resulting in 2048-dimension output vector for an arbitrary-length input audio. The pooled vector goes to two fully-connected layers with ReLU activation on the first to output 527 classes. The pooling layer aggregates information across all time steps into a fix-dimension vector, and it gives a single prediction output for the entire 10s training file. While the pooling layer gives good results for the weak label training objective, we learned that an embedding trained this way has limitations in FSL of temporal sequences.

\textbf{Knowledge Distilled Embedding}. For on-device inference, it is desirable to use a model that has good performance versus complexity trade-offs.  We modify the MobileNetV2 \protect\cite{sandler2018mobilenetv2} such that the channel dimension of the last convolutional layer is changed from 1280 to 512, named as MobileNetV2-S. Then we adopt Knowleddge Distillation (KD) \protect\cite{hinton2015kd} to further improve its performance. A larger MobileNetV2-L with a 1.4x width multiplier applied to each layer of MobileNetV2-S and 1,500 channels in last convolutional layer is used as the teacher. The student MobileNetV2-S-KD is trained by reducing both K-L divergence between teacher-student logits and student categorical cross entropy loss. Despite being $\sim$1/13 the size of the reference model, it achieves good performance (see Section \protect\ref{sec:results}). The detector design is similar, aside from having different embedding dimensions and the number of FC layers reduced to 1.

\textbf{Strong Label Embedding}. While the pooling layer in two previous architectures provide a way to train with weakly labeled datasets, it has the potential downside of not preserving temporal dynamics within the audio.  If a strongly labeled dataset were available, we could create an output layer that makes frame-level predictions.  In the strong label embedding, we preserve the same MobileNetV2-S backbone, but we pool across the frequency dimension but not time, thus preserving the temporal dimension. The output of the pooling goes into a 1x1 Conv1d, which produces 527 class outputs every 320ms. This model is named as MobileNetV2-S-Strong and its training is based on MobileNetV2-S-KD initialization. 

\subsection{Model training}

\textbf{Weak label training}. We adopt various known audio classification training techniques from previous work \protect\cite{huang2018aclnet}, \protect\cite{kong2020panns}: 1) random resampling the audio in the +/-10\% range, 2) SpecAugment \protect\cite{park2019specaugment}, 3) mixup \protect\cite{zhang2017mixup}, 4) +/-20 dB random gain, and 5) class imbalanced sampling in data loader to boost the occurrences of minority classes.  We use the AdamW optimizer \protect\cite{loshchilov2018fixing} with a one-cycle learning rate schedule that warms up to learning rate of 0.01, then decaying to 0.0001 at the last epoch 30.  With these settings we achieve satisfactory results for the Resnet teacher model.  The weak label embedding was trained with a nearly identical pipeline, except it used an additional training target from the teacher model output.

\textbf{Pseudo-strong label training}. Frame-level training targets are used for training the strong label embedding.  However, performing strong label is a labor-intensive, time-consuming and expensive task.  Instead, we resort to utilizing the weak label reference model for producing pseudo-strong labels.  Empirically we found that the models trained with weak labels can produce reasonably good predictions when input duration is much shorter than the 10s training samples.  For every 10 s AudioSet file, we generate ResNet model predictions at 100 ms hops, with 0.5s input duration.  We apply a threshold of 0.5 to make each class output binary.  The pseudo labels are used for training the strong label embedding.  Except for training targets, the training pipeline is the same as in weak label training.

\section{Application: Few Shot Detection}
\label{sec:few-shot-detection}

\textbf{Audio Curation}.
Instead of assuming that few shot sound events are already segmented, we use the embedding to discover target sounds and their boundaries. This allows for few-shot learning from audio without onset or offset annotations.

Given $K$ shots of arbitrary lengths each with one instance of the target sound, we need to estimate a set of onset and offset timestamps $D_{enroll} = \{(\mathbf{x}_{i},y_{i})\}_{i=1}^{K}$ without any prior information about the target. We approached this as follows:
\begin{enumerate}
  \item Find loud segments within each audio shot. A logistic regression model trained with $loud$ labels (top 5 percentile of logmel frames with highest energy) and $quiet$ (bottom 5 percentile) is used to find the segments classified as $loud$, for each shot independently.
  \item Find segments that semantically similar, in the acoustic representation space, and are present across shots. Cosine distance of the fixed length embedding between pairs is used to group them, and select only the first onset and last offset on each shot, discarding the rest.
  \item Using the shortest of the segments as an exemplar, the rest of segments are adjusted to have the same length by finding the subsegment with exemplar length that gives the highest cross-correlation in the logmel domain. 
\end{enumerate}

\textbf{Binary Classifier Training Method}.
Given strong-label dataset $D_{enroll}$, our goal is to produce the augmented dataset \\ $D_{train} = \{(e(\mathbf{x}_{i}),y_{i})\}_{i=1}^{K+A}$, where $K >> A$. This requires increasing intra-class variance to approximate aspects of inference-time conditions known to be lacking in the available set of target samples, by augmenting positives and generating negatives. We then use augmented $D_{train} = \{(e(\mathbf{x}_{i}),y_{i})\}_{i=1}^{K+A}$ for training a low footprint, generalizable binary classifier. To do so, we first extract pretrained variable length embedding sequences as features (from segmented audio), using pretrained embedding $e(\cdot)$.

\textit{Target Class Embedding Augmentation}.
We seek to augment the target class. We approach this in the time-domain, by slightly enlarging the segmentation output of audio curation (onset, offset), by 500ms, and time-shifting within the enlarged bounds. We also augment in the embedding domain, by adapting the $\Delta$-encoder \protect\cite{schwartz2018delta} to the time-domain, by treating 1024-d embedding sequences as sets of fixed length embeddings. Deformations are learned from pairs of embedding frames. Embedding pairs are assumed to come from pairs of sound events of the form $X$, $\tilde{X}$. Thus, learned deformation $z_i$ is interpreted as the embedding-domain operation which synthesized acoustically degraded $e_T{\tilde{X}}$ from paired clean sample $e_T{X}$: the embedding-domain analog of convolving a clean sound event sample with a room impulse response (RIR). 

\textit{Non-Target Class Embedding Synthesis}.
To source negative examples, we apply parametric, embedding-domain masking and shuffling operations to augmented target sequences. In masking an appropriate length of contiguous embedding frames, or shuffling blocks of appropriate length (both conditional on length of target sequence), we minimally perturb target audio representations. Negatives are thus synthesized to lie close to the ultimate decision boundary.

\textit{Learning}.
Finally, we propose an approach to train a small-footprint model that learns a generalizable decision boundary. We implement a learnable linear projection for dimensionality reduction, temporal modeling via 1-d dilated causal convolutions, and a binary classifier in a single end-to-end model, and train it via a multi-loss objective which enforces a large-margin loss \textit{w.r.t.} the input and intermediate feature maps. The formulation of distance to decision boundary we employ (\protect\ref{eq4}) is a modification of a previously proposed large margin loss \protect\cite{elsayed2018margin}. In conjunction with a small term for binary cross entropy, this multi-objective loss enables improved learning on longer sequences.

\vspace{-2mm}
\begin{equation}
\begin{aligned}
{\tilde{d_n} = \frac{ f_i(x_n) - f_{\neg i}(x_n) }{ \| h_{T}[\nabla_{x_n} f_i(x_n) - \nabla_{x_n} f_{\neg i}(x_n)] \|_{2}^{F} + \epsilon }}, \\
\end{aligned}
 \label{eq4}
\end{equation}

\vspace{-2mm}
\section{Experiments}
\label{sec:results}
\vspace{-2mm}

We employ AudioSet and an internally collected dataset of naturalistic domestic sound events and scenes for evaluation. 
Specific acoustic sequences are sourced from this latter dataset, due to the scarcity of finer-grained subclass labels relevant to sound events classes which have distinctive temporality in typical annotated sound classification corpora. All datasets are resampled to 16 kHz.

\begin{table}[th]
\centering
\caption{Acoustic representation pretraining approaches: \\ In-domain evaluation (Audioset) }
\label{tab:pretrained_performance}
  \begin{tabular}{lcccccc}
    \toprule
    \multirow{1}{*}{System} &
      \multicolumn{1}{c}{\textbf{Params}} &
      \multicolumn{1}{c}{\textbf{mAP}} &
      \multicolumn{1}{c}{\textbf{d-prime}} \\
      \midrule
    \label{tab:results}
    Reference                     & 25.6M   & 0.439 & 2.72  \\
    MobileNetV2-L                 & 4.45M & 0.394 & 2.63  \\
    MobileNetV2-S                 & 1.98M & 0.376 & 2.53  \\
    MobileNetV2-S-KD              & 1.98M & 0.401 & 2.67  \\
    MobileNetV2-S-Strong & 1.98M & 0.254 & 2.43  \\
    \bottomrule
  \end{tabular}
\end{table}

\textbf{Pretrained Model Benchmarking}. 
In Table \protect\ref{tab:pretrained_performance}, we report sound classification results of various embeddings. First, the reference model achieves 0.439 mAP, outperforming the 14-layer PANN \protect\cite{kong2020panns}. Second, our distillation based MobileNetV2-S-KD achieves a much smaller footprint than the reference model and 2.2x model compression from the MobileNetV2-L teacher network. Surprisingly, the student (0.401 mAP) outperforms the teacher (0.394 mAP). Third, MobileNetV2-S without distillation achieves 0.376 mAP. It indicates the importance of extracting information from the teacher. Finally, we show mAP and d-prime of the fine temporal resolution strong-label embedding MobileNetV2-S-Strong. Note that this model employs 100 ms frame-level pseudo-strong labels generated by the reference model predictions. Labels estimated in this way can diverge from the weak-labeled ground truth. However, because of the desirable fine temporal details learned by the strong-label embedding, we still adopt MobileNetV2-S-Strong for few-shot detection and results in the next subsection show its advantages. 

\textbf{Analysis of Proposed Framework}. 
We evaluate fine-grained robustness in an out-of-distribution (OOD) detection task involving two practical assumptions: that (a) within-class variance of positive samples fails to match inference-time conditions, and (b) no assumed access to non-target samples. 

Each inference-time task was comprised of of three target samples and one hundred and eighty non-target samples, corresponding to the one target sequence against nine unseen non-target sequences. Recordings of each sequence in the evaluation set were split 50\% near-field and 50\% far-field. Acoustic content of non-target samples varied across evaluation tasks, in correspondence with the domestic environment they were recorded in. 

\begin{figure}[!t]
\begin{minipage}[b]{1.0\linewidth}
  \centering
  \centerline{\includegraphics[width=9.0cm]{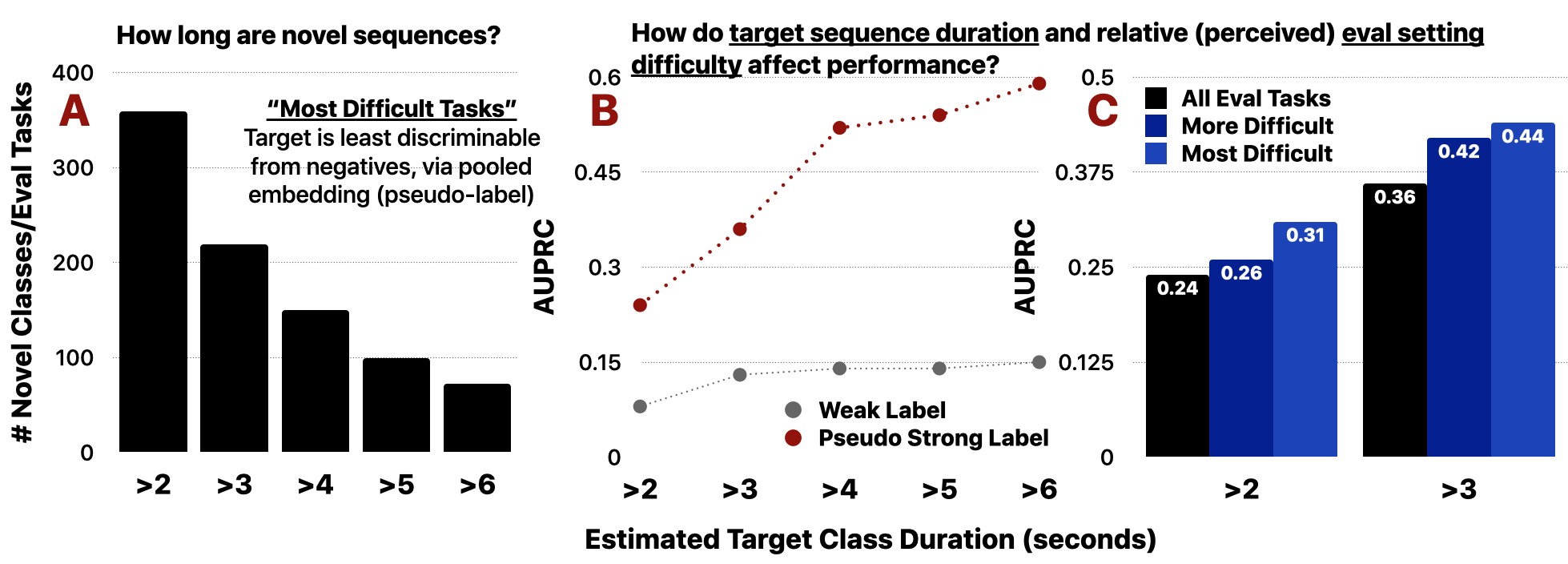}}
\end{minipage}
\caption{(A) Duration distribution of novel sequences evaluated (estimated via curation). (B) Longer duration target sequences achieve larger relative performance improvements (metric: median of avg. AUPRC over 10 reps/enrollment sequence). (C) Target length vs. Evaluation setting difficulty.}
\label{fig:fig3}
\end{figure}

Fig. \protect\ref{fig:fig3}(A) shows the data distribution of evaluated target sequences. Fig. \protect\ref{fig:fig3}(B) shows the performance of the proposed few shot approach, with pseudo-strong labeled embedding (PSL) outperforming the corresponding weak labeled embedding (WL) overall: As the duration of target sequence increases, we observed a greater relative improvement of the PSL-based few shot system. This suggests the benefit of the proposed pretrained embedding alongside the proposed system, for the task of longer-duration acoustic sequence detection tasks which have been explored to a very limited extent in the literature thus far. 
Finally, relative difficulty across tasks is measured via a ``task difficulty index": a scalar difficulty measure computed via the cosine distance across fixed-length embeddings. As shown in Fig. \protect\ref{fig:fig3}(C), as the evaluation task gets harder, the relative benefit of the proposed system is elevated. This suggests a potential complimentarity between our proposed method and fixed-length embedding FSL methods explored more widely. In future work, we plan to investigate the complimentarity of such methods.

\vspace{-3mm}
\section{Conclusion}
\label{sec:conclusion}
\vspace{-2mm}

In this work we proposed a unified approach for few shot detection of novel acoustic sequences, suitable for resource-constrained use cases. Our study consists of methods for acoustic representation pretraining and FSL system components designed on top. Our proposed embedding pretraining, distillation, and pseudo-strong-labeling benchmark favorably against current approaches. We used pretrained audio representations for few shot learning of acoustic sequences, and found proposed representations to enable few shot SED for underexplored tasks, evaluated in challenging conditions.

\clearpage

\bibliographystyle{IEEEbib}
\bibliography{ms}

\end{document}